\documentclass[%
preprint,
nofootinbib,
nobibnotes,
 amsmath,amssymb,
 aps,
 prd,
]{revtex4-2}

\usepackage[utf8]{inputenc}
\usepackage[T1]{fontenc}
\usepackage{natbib}
\usepackage{slashed}
\usepackage{graphicx}% Include figure files
\usepackage{dcolumn}% Align table columns on decimal point
\usepackage{bm}% bold math
\usepackage{aas_macros}
\usepackage{color}
\usepackage{physics}
\usepackage{booktabs}
\usepackage{comment}
\usepackage{xspace}   % xspace for new command definition
\usepackage{url} % url typesetting
\usepackage{siunitx}
\usepackage{hyperref} % hypetextual links
\usepackage{xurl} % allows line breaking in urls (to be loaded after bibliography packages, hyperref and url)
\allowdisplaybreaks
\newcommand{\ped}[1]{_{\scriptscriptstyle{\textrm{#1}}}}

\newcommand{\peff}{p\ped{eff}}

\newcommand{\Weff}{W\ped{eff}}

\newcommand{\MadDM}{\protect{\tt MadDM}\xspace}
\newcommand{\DarkSUSY}{\protect{\tt DarkSUSY}\xspace}
\newcommand{\MicrOMEGAs}{\protect{\tt MicrOMEGAs}\xspace}

\newcommand{\SuperIsoR}{\protect{\tt SuperIso Relic}\xspace}

\newcommand{\DarkPack}{\protect{\tt DarkPACK}\xspace}

\newcommand{\sigmav}{\expval{\sigma v}}

\newcommand{\Ho}{H_0}

\newcommand{\Omm}{\Omega\ped{m}}
\newcommand{\Omb}{\Omega\ped{b}}
\newcommand{\Omc}{\Omega\ped{c}}

\newcommand{\Oml}{\Omega_{\Lambda}}

\newcommand{\vmol}{{v_{\scriptscriptstyle\textrm{M\o l}}}}

\newcommand{\sigmavmol}{\langle \sigma \vmol \rangle}

\newcommand{\sveff}{\sigmavmol\ped{eff}}

\newcommand{\ee}{\mathrm{e}}
\newcommand{\ii}{\mathrm{i}}

\definecolor{bg}{rgb}{0.95,0.95,0.95}

\title{Revisiting the averaged annihilation rate of thermal relics at low temperature}

\begin{document}

\hfill {\tt CERN-TH-2023-232}

\def\thefootnote{\fnsymbol{footnote}}

\begin{center}
{\Large
\vspace{1cm}
Revisiting the averaged annihilation rate of thermal relics\\ at low temperature}

\vspace{1.5cm}
{\large\bf  
A.~Arbey$^{a,b,}$\footnote{Email: alexandre.arbey@ens-lyon.fr},
F.~Mahmoudi$^{a,b,c,}$\footnote{Email: nazila@cern.ch},
M.~Palmiotto$^{a,}$\footnote{Email: marco.palmiotto@univ-lyon1.fr}
}
 
\vspace{0.5cm}
{\em $^a$Universit\'e de Lyon, Universit\'e Claude Bernard Lyon 1, CNRS/IN2P3, \\
Institut de Physique des 2 Infinis de Lyon, UMR 5822, \\F-69622, Villeurbanne, France}\\[0.2cm]
{\em $^b$Theoretical Physics Department, CERN, CH-1211 Geneva 23, Switzerland}\\[0.2cm] 
{\em $^c$Institut Universitaire de France (IUF)}
\end{center}

\renewcommand{\thefootnote}{\arabic{footnote}}
\setcounter{footnote}{0}

\vspace{1cm}

\vspace{1.cm}
\thispagestyle{empty}
\centerline{\bf ABSTRACT}
\vspace{0.5cm}

We derive a low-temperature expansion of the formula to compute the average annihilation 
rate $\sigmav$ for dark matter in $\mathbb{Z}_2$-symmetric models, both in the absence and the presence of mass degeneracy in the spectrum near the dark matter candidate.
We show that the result obtained in the absence of mass degeneracy is compatible with the analytic formulae in the literature, and that it has a better numerical behaviour for low temperatures.
We also provide as ancillary files two {\tt Wolfram Mathematica} notebooks which perform the two expansions at any order.

\newpage

\section{Introduction}

%\subsection{Importance of dark matter in astroparticle physics}
One of the main areas of research in astroparticle physics today is the search for dark matter (DM).
For decades, a number of astrophysical observations have been impossible to explain in the context of general relativity (GR), if we also assume that the Standard Model (SM) of fundamental interactions describes the entire particle content of the universe.
Hence, a hypothesis that can explain most - or in some contexts all - observations is the existence of a kind of stable and non-relativistic matter that couples very weakly with SM fields.
For this reason, this kind of matter is referred to as \emph{dark matter}.
Aside from the astrophysical observations, the existence of DM is also a necessity in cosmology in order to obtain a coherent description of the growth of perturbations. 

Moreover, despite the fact that it describes very well the phenomena observed up to the TeV scale,\footnote{Except for neutrinos' flavour oscillations.}
the SM is expected to fail at a certain energy scale, surely lower than the Planck energy.
Thus, if DM is in part composed of stable particles, it could be explained in some extensions of the SM. Alternatively, it is possible to extend the SM with the aim of having a model that describes the nature of a fraction - or the totality - of the DM abundance.

%\subsection{DM abundance and relic density}
The abundance of DM has been measured by the Planck collaboration~\cite{Planck:2018vyg} as the 
relative energy density:
\begin{align}
 H_0 &={67.66\pm0.42}\,\si{km/s/Mpc},\\
 \Omm  &= 0.3111 \pm 0.0056,\\
 \Omb h^2 &= 0.02237 \pm 0.00015, \\
 \Omc h^2 &= 0.1200 \pm  0.0012, \\
 \Oml &= 0.6847 \pm 0.0073,
\end{align}
where $H_0$ is the Hubble parameter, $h$ is the reduced Hubble parameter, defined as:
\begin{align}
 h = \frac{\Ho}{\SI{100}{km/s/Mpc}},
\end{align}
and where $\Omb$ is the relative density of \emph{baryonic} matter, i.e.~of the fraction of cold matter visible via electromagnetic signals, and $\Omc$ is the relative density of \emph{cold dark matter}, i.e.~of the 
fraction of cold matter electromagnetically invisible.
The discrepancy between $\Omm$ and $\Omb$ is the first evidence of the existence of cold dark matter, and that this component dominates the non-relativistic matter content of the Universe.

%\subsection{Accuracy of the calculation}
The contribution to the relative density of a particle's species can be computed by solving the 
Boltzmann equation for its number density, then obtaining from it the energy density, and finally 
dividing this result by the critical density of the Universe today.
The form of the Boltzmann equation and its resolution for the non-relativistic case
can be found e.g.~in \cite{KolbTurner:1990}. In particular, it has the form
\begin{align}\label{eqn:Boltzmann1}
    \dv{n}{t} = -3Hn + \expval{\sigma v}(n - n\ped{eq}),
\end{align}
where $n$ is the number density of the dark matter candidate, as a function of the time $t$,\footnote{In the
Friedmann-Lama\^itre-Robertson-Walker metric.}
$n\ped{eq}$ is the value of equilibrium for $n$, at the temperature $T$ corresponding to the time $t$, 
and $\expval{\sigma v}$ is the thermal average of the product of the total cross-section of annihilation
of dark matter candidates into SM particles with the relative velocity of the particles in the initial state.
In order to compute the thermal average, it has been assumed that the particles in the initial state 
have non-relativistic velocities, allowing the replacement $s \to 4m^2 + m^2v^2$, where $s$ is the Mandelstam
variable, $m$ is the mass of the DM candidate, and $v$ its velocity in the centre-of-mass frame. 
The expanded $\sigma v$ is then averaged, yielding an expansion in $x=m/T$ for  $\expval{\sigma v}$.
The work by Srednicki et al.~\cite{Srednicki:1988ce} aimed then to have a more reliable 
expansion of $\expval{\sigma v}$, by finding the general non-relativistic formula expressed
directly in powers of $1/x$, and starting from the squared matrix elements of the annihilation reactions.

In the context of DM produced via annihilation and co-annihilation, 
the work of Gondolo and Gelmini \cite{Gondolo:1990dk}, and of Edsjo and Gondolo \cite{Gondolo:1998ah}
made a step forward, in generalising the equation in the relativistic case. 
Firstly, it is pointed out that $v$ should not be the relative velocity, but the M\" oller velocity, thus
making $\expval{\sigma v}$ a scalar, from now on denoted as $\sveff$.
Then, the scenario of annihilation to SM particles and of co-annihilation is considered in the models 
with a $\mathbb{Z}_2$ symmetry that prevents the DM candidate from decaying into SM particles. 
The result is a Boltzmann equation for the total number density of the species with the same 
$\mathbb{Z}_2$-parity as the DM candidate, which has the same form as equation \eqref{eqn:Boltzmann1}.
In this context, the expression of $\sveff$ is derived and linked to the one
already presented in \cite{Srednicki:1988ce}.

In this work, we re-consider the formula derived in \cite{Gondolo:1998ah} for $\sveff$ 
in \autoref{ssec:relic}, also showing why the implementation of such a formula 
can lead to numerically unreliable results at low temperatures.
In \autoref{ssec:lowtbody}, we point out that a numerical evaluation of such a formula at low 
values for $T$ presents some numerical issues, and we derive its expansion in $1/x$, by 
following the procedure outlined in \cite{Cannoni:2013bza} by Cannoni. 
Finally, in \autoref{ssec:lowtbodysplit} we generalise the expansion in the case of small mass splitting 
in the spectrum near the DM candidate's mass.
We conclude in \autoref{ssec:conclusion} by discussing the results and their areas of application.

\section{Freeze-out scenario for thermal relic density}\label{ssec:relic}

The standard scenarios for dark matter particles are the so-called thermal relic scenarios, in which a single relic particle can explain the nature of dark matter. In the freeze-out scenarios, the new physics particles are considered in thermal equilibrium at a common temperature $T$. The expansion rate $H$ of the Universe is given by the Friedmann equation:
\begin{equation}
H^2=\frac{8 \pi G}{3} g\ped{eff}(T) \frac{\pi^2}{30} T^4\,,\label{eq:friedmann_stand}
\end{equation}
where $g\ped{eff}$ is the effective number of degrees of freedom of radiation.

At thermal equilibrium, under the assumption of the Maxwell-Boltzmann statistics, the total number density of new physics particles is given by
\begin{equation}
n\ped{eq} = \frac{T}{2\pi^2} \sum_i g_i m_i^2 K_2\left(\frac{m_i}{T}\right)\,,
\end{equation}
where $g_i$ and $m_i$ are the number of degrees of freedom and the mass of  the $i$-th new physics particle, respectively, and $K_2$ the modified Bessel function of the second kind of order 2.

To compute the present relic density of dark matter particles, one needs to solve the Boltzmann evolution equation \cite{Gondolo:1990dk,Edsjo:1997bg,Arbey:2021gdg}:
\begin{equation}
\dv{n}{t}=-3Hn-\langle \sigma\ped{eff} v\rangle (n^2 - n\ped{eq}^2)\,, \label{eq:evol_eq}
\end{equation}
where $n$ is the total number density of new physics particles and $\langle \sigma\ped{eff} v\rangle$ is the thermal average of the annihilation rate of the new physics particles to the Standard Model particles. 

The thermal average of the effective cross-section at temperature $T$ is obtained, under the assumptions of thermal equilibrium and Maxwell-Boltzmann statistics:
\begin{equation}
\langle \sigma\ped{eff}v \rangle (T) = \dfrac{\displaystyle\int_0^\infty \dd p\ped{eff} p\ped{eff}^2 W_{\rm{eff}}(\sqrt{s}) K_1 \left(\dfrac{\sqrt{s}}{T} \right) } { m\ped{DM}^4 T \left[ \displaystyle\sum_i \dfrac{g_i}{g\ped{DM}} \dfrac{m_i^2}{m\ped{DM}^2} K_2 \left(\dfrac{m_i}{T}\right) \right]^2}\,,\label{eqn:sigmavefffin2}
\end{equation}
where $K_1$ is the modified Bessel function of the second kind of order 1, $g\ped{DM}$ and $m\ped{DM}$ are the number of degrees of freedom and the mass of the dark matter particle, and
\begin{equation}
p\ped{eff}(\sqrt{s}) = \frac{1}{2} \sqrt{s -4 m\ped{DM}^2} \,,
\end{equation}
where $\sqrt{s}$ is the centre-of-mass energy. We can obtain $W_{\rm eff}$ by integrating over the outgoing directions of the final particles \cite{Arbey:2021gdg}:
\begin{equation}
\dv{\Weff}{\cos(\theta)}= \sum_{ijkl} \frac{p_{ij} p_{kl}}{ 8 \pi g\ped{DM}^2 p_{\rm eff} S_{kl} \sqrt{s} }
\sum_{\rm helicities} \left| \sum_{\rm diagrams}  \mathcal{M}(ij \to kl) \right|^2 \,,\label{dWeff}
\end{equation}
where $\mathcal{M}(ij \to kl)$ is the amplitude of two new physics particles $(i,j)$ giving two Standard Model particles $(k,l)$, and $\theta$ is the angle between particles $i$ and $k$, $S_{kl}$ is a symmetry factor equal to 2 for identical final particles and to 1 otherwise, and $p_{kl}$ is the final centre-of-mass momentum such that
\begin{equation}
p_{kl} = \frac{\left[s-(m_k+m_l)^2\right]^{1/2} \left[s-(m_k-m_l)^2\right]^{1/2}}{2\sqrt{s}}\,.
\end{equation}

The current density of dark matter particles can be obtained by integrating the Boltzmann equation (\ref{eq:evol_eq}) between a high temperature where all particles are in thermal equilibrium, and the current Universe temperature $T_0=2.726$ K. The freeze-out temperature $T_f$ is defined as the temperature at which the dark matter particles leave thermal equilibrium. 

There exist several codes for the calculation of dark matter relic density, such as \SuperIsoR\cite{Arbey:2009gu,Arbey:2011zz,Arbey:2018msw}, \MicrOMEGAs \cite{Belanger:2001fz,Belanger:2018ccd}, \DarkSUSY \cite{Gondolo:2004sc,Bringmann:2018lay}, \MadDM \cite{Ambrogi:2018jqj,Arina:2020kko} and \DarkPack \cite{Palmiotto:2022rvw}, which use different methods of integration of the Boltzmann equation and calculation of the thermal average of the effective cross-section.

In particular, one can observe that the formula \eqref{eqn:sigmavefffin2} can present some numerical instabilities 
for small values of $T$. In fact, both the Bessel functions have an asymptotic behaviour, as their
expansion is given in \eqref{eqn:asympKn}, so for large $1/T$ both the integrand function in the numerator and 
the sum in the denominator tend to 0, leading to an undefined form. Thus, when the evaluation of the numerator
or the denominator returns a number close to the minimum value of the adopted floating number precision, 
the value of $\sigmav$ cannot be reliable.

\section{Averaged annihilation rate at low temperature}
\label{ssec:lowtbody}

The definition of the averaged annihilation rate given in Eq.~\eqref{eqn:sigmavefffin2}, is the central part of the Boltzmann equation. In fact, for small values of $T$, the arguments of the Bessel functions tend to infinity, and 
both $K_1$ and $K_2$ become infinitesimally small since their arguments tend to infinity.
This generates some computational issues, if $T$ is very small, which is the case in the recent Universe. 

From a phenomenological perspective, often the freeze-out temperature will not be small enough to require a specific expansion for $\sveff$. In fact, it is typically equal to the mass of the DM candidate times a factor ranging from 1/30 to 1/20, and therefore there is no need to evaluate $\sveff$ at 
temperatures as low as $\SI{e-14}{\giga\electronvolt}$.
However, we found this expansion useful, since in some cases it is possible to calculate, or to 
find in the literature, some formulae for $\sveff$ in the non-relativistic case. Thus, providing 
a correct numerical expansion at low temperature, independent from a full formula prone to numerical 
instabilities, allows us to detect possible errors in the numerical implementation of the model, or in the 
derivation of an analytical expression of the non-relativistic $\sveff$ in a specific model.

It is therefore useful to study the expansion of the averaged annihilation rate at low temperatures,
in order for example to verify that the relativistic result is consistent with the non-relativistic one. The derivation of the latter can be found in Ref.~\cite{Srednicki:1988ce}. 

In this subsection, we will outline the steps of the expansion of \eqref{eqn:sigmavefffin2}, showing that it can be performed
up to any given order.
We also show that the lowest order is the order zero, hence proving that the formula \eqref{eqn:sigmavefffin2} does not present singularities at $T=0$.
The original procedure has been suggested in Ref.~\cite{Cannoni:2013bza}, and in the following we describe the final computational steps in a way that they can be reproduced by hand or even with symbolic manipulation algorithms. We also provide as an ancillary file a {\tt Mathematica notebook} \cite{mathematica} which performs such an expansion.

To begin, we make a change of variable for the integral \eqref{eqn:sigmavefffin2}:
\begin{align}
 \peff \to y = \frac{\peff^2}{m_{1}^2} +1,
\end{align}
where $m_{1}$ is the mass of the lightest new physics particle, i.e.~the dark matter particle, which we will denote in the following $\chi_1$. In the denominator, we keep in the sum only the contribution of $\chi_1$, since it is the lightest particle leading to the dominant contribution to the sum. Using the asymptotic form of $K_n$ provided in the Appendix in Eq.~\eqref{eqn:asympKn}, we therefore obtain:
\begin{align}
 \sveff = \frac{x}{2m_1^2K_2(x)}\int_1^{+\infty} \dd{y} \sqrt{y-1}\Weff(y)K_1\qty(2x\sqrt{y}),
\end{align}
where $x= m_1/T$.

Similarly to $K_2$, $K_1$ has its maximum value when its argument has its smallest value in the integral. This means that the largest contributions to the integral are coming from the region with $y \gtrsim 1$. Let us then expand $\Weff$ around $y=1$:
\begin{align}
 \Weff(y) = \sum_{n=0}^\infty \frac{1}{n!} (y-1)^n W_n,
\end{align}
where $W_n =\displaystyle \eval{\dv[n]{\Weff}{y}}_{y=1}$. Then

\begin{align}
\sveff = 
\sum\limits_{n=0}^{\infty} \frac{1}{n!} W_n
\mathcal{K}_n (x),
\label{eqn:svexpansion}
\end{align}
where we have defined:
\begin{align}\label{eqn:kandI}
\mathcal{K}_n (x)&=\frac{x}{2 m_1^2 K^2_2(x)} I_n (x), &
I_n (x)&=
\int_{1}^{+\infty}\dd{y}
(y-1)^{\frac{1}{2} +n}  K_1 (2x\sqrt{y}).
\end{align}

The integral in $I_n$ corresponds to Eq.~(\ref{I_GR}) with $\lambda=0$, $\mu=3/2+n$ and $\nu=1$. Hence,
\begin{equation}
I_n (x)
=\frac{1}{2} \Gamma \left(n+\frac{3}{2}\right) 
G_{1,3}^{3,0}
\begin{pmatrix}
x^2 \Biggr\rvert
\begin{array}{c}
 0 \\
\displaystyle-n-\frac{3}{2}, \frac{1}{2}, -\frac{1}{2}
\end{array}
\end{pmatrix}.
\end{equation}
The coefficients $\mathcal{K}_n (x)$ of expansion \eqref{eqn:svexpansion} are therefore
\begin{flalign}
\mathcal{K}_n (x)=
\Gamma\qty(n+\frac{3}{2}) 
\frac{x}{4 m_1^2 K^2_2(x)}
G_{1,3}^{3,0}
\begin{pmatrix}
x^2 \Biggr\rvert
\begin{array}{c}
 0 \\
\displaystyle -\frac{1}{2},\frac{1}{2},-n-\frac{3}{2}
\end{array}
\end{pmatrix}.
\label{eqn:expansion_rho=0}
\end{flalign}
Using the results in Appendix~\ref{ssec:Meijer}, we can write the asymptotic form of 
$\mathcal{K}_n (x)$ as: 
\begin{align}\label{eqn:asympkn}
\mathcal{K}_n (x)=
\Gamma\qty(n+\frac{3}{2}) \frac{\sqrt{\pi}}{4 m_1^2}\ee^{-2x}
\frac{x}{ K^2_2(x)}\sum_{p = n+2}^\infty g_{n,p}x^{-p}.
\end{align}
We consider now the asymptotic form of $K_2(x)$. By using Eq.~\eqref{eqn:asympK22n}, 
and keeping the same notation as in Appendix~\ref{ssec:Bessel}, we can write:
\begin{align}
\mathcal{K}_n (x)=
\Gamma\qty(n+\frac{3}{2}) \frac{1}{2 \sqrt{\pi} m_1^2}
\frac{x^2}{1+B(x)}\sum_{p = n+2}^\infty g_{n,p}x^{-p}.
\end{align}
Since $B(x)$ is a small quantity, we can use the properties of the geometric sum and write:
\begin{align}
\mathcal{K}_n (x)=
\Gamma\qty(n+\frac{3}{2}) \frac{1}{2 \sqrt{\pi} m_1^2}
{x^2}\sum_{r=0}^{\infty} (-1)^r B^r(x)\sum_{p = n+2}^\infty g_{n,p}x^{-p}.
\end{align}
At this point, by knowing the $g_{n,p}$, the procedure becomes straightforward.

First, we use the expansions \eqref{eqn:svexpansion} and  
\eqref{eqn:asympkn} to factorise the terms independent of $n$ in the expression of $\sveff$:
\begin{align}\label{eqn:sveffseries0}
 \sveff &= \frac{1}{2 \sqrt{\pi} m_1^2}
{x^2}\sum_{r=0}^{\infty} (-1)^r B^r(x) \sum\limits_{n=0}^{\infty} \qty[\frac{1}{n!} W_n\Gamma\qty(n+\frac{3}{2})\sum_{p = n+2}^\infty g_{n,p}x^{-p}]\,.
\end{align}
The term in square brackets is a Laurent series whose maximum order is 2. So we can 
define a set of $g_p$ such that:\footnote{Note that for the sum over $p$ we kept the same name for the index for clarity. In fact, the powers of $x$ are expressed as functions of $p$ in the original sum. This means that if truncate at a certain order, the upper limit of the sums is the same.}
\begin{align}\label{eqn:sveffseries1}
 \sveff &=  \frac{1}{2 \sqrt{\pi} m_1^2}
{x^2}\sum_{r=0}^{\infty} (-1)^r B^r(x) \sum\limits_{p=2}^{\infty} \frac{g_p}{x^p}\,.
\end{align}
Moreover, using Eq.~\eqref{eqn:BBtildeseries} we can define the coefficients $\beta_r$ such that
\begin{align}
  \sveff &=  \frac{1}{2 \sqrt{\pi} m_1^2} \sum_{r=0}^{\infty} \frac{\beta_r}{x^r} \sum\limits_{p=0}^{\infty} \frac{\tilde g_p}{x^{p}}\,,
\end{align}
where $\tilde g_p = g_{p+2}$.
Written in this form, it is clear that the lowest order is zero, as it should be.\footnote{Note that for the sum over $r$ we kept the 
same name for the index for clarity. $B^r$ gives the highest contribution to the term $\eta_r/x^r$. Therefore, if we truncate 
at a certain order, the range of the two sums is the same.}

The next step is to determine the coefficients of the powers of $1/x$ up to a given order $N\ped{max}$.
This can be done once we know the coefficients $\beta_r$ and $g_p$ up to $r = N\ped{max}$ and $p = N\ped{max} +2$.
We can also show that there is a maximum contribution from $n$, which can be obtained from the range of the sum in $p$:
\begin{align}
 n+2 \leq p \leq N\ped{max} +2
\end{align}
from which we obtain the condition $n\leq N\ped{max}$.
To summarise, in order to truncate the expansion at the order $N\ped{max}$, the indexes have the following 
ranges:
\begin{align}\label{eqn:ordersexpansion}
 0 \leq &r \leq N\ped{max}, & 0 \leq &n \leq N\ped{max}, & n+2 \leq &p \leq N\ped{max} +2\,.
\end{align}

Let us now how discuss to perform the expansion, considering for instance $N\ped{max}=4$ to illustrate the intermediate steps and $N\ped{max}=10$ for the final result.
The {\tt Mathematica notebook} provided as an ancillary file provides the algorithm valid for any values of $N\ped{max}$.

For a given $N\ped{max}$ the maximum order of the derivative of $\Weff$ that contributes 
to $\sveff$ is exactly $N\ped{max}$:
\begin{align}
\sveff = 
\sum\limits_{n=0}^{N\ped{max}} \frac{1}{n!} W_n
\mathcal{K}_n (x) \,.
\end{align}
$\sveff$ is defined as a finite sum, and each $\mathcal{K}_n$ is the product of two series that we know where to truncate. Let us define the quantity:
\begin{align}
 D(x)= \frac{1}{1+B(x)} = \sum_{r=0}^{N\ped{max}} (-1)^r B^r(x)\,.
\end{align}
Then, we can write $\mathcal{K}_n$ in the form:
\begin{align}\label{eqn:kappanseries}
\mathcal{K}_n (x)=
\Gamma\qty(n+\frac{3}{2}) \frac{1}{2 \sqrt{\pi} m_1^2}
{x^2}D(x)\sum_{p = n+2}^{N\ped{max}+2} g_{n,p}x^{-p}\,.
\end{align}
The coefficients $g_{n,p}$ are tabulated in Eq.~\eqref{eqn:gnpvalues}.
Therefore, we are left with determining the coefficients of the expansion of $D$ and calculating the product of the two truncated series. Firstly, we write each power of $B$ in $D$ by expanding the series of $B$ in $B^r$ up to the order $N\ped{max}-r+1$. Then, up to the 4th order, the non-trivial powers of $B$ are:
\begin{align}
 B^2 &=  \frac{b_2^2+2 b_1 b_3}{x^4}+\frac{2 b_2 b_1}{x^3}+\frac{b_1^2}{x^2},\\
 -B^3 &= -\frac{3 b_2 b_1^2}{x^4}-\frac{b_1^3}{x^3},\\
 B^4 &= \frac{b_1^4}{x^4}.
\end{align}
The coefficients $b_i$ are given in \autoref{tab:abbtilde}, and the coefficients $\beta_r$ of the expansion of $D(x)$ 
are given in \autoref{tab:betar}.

\newpage

\begin{figure}\centering
\includegraphics{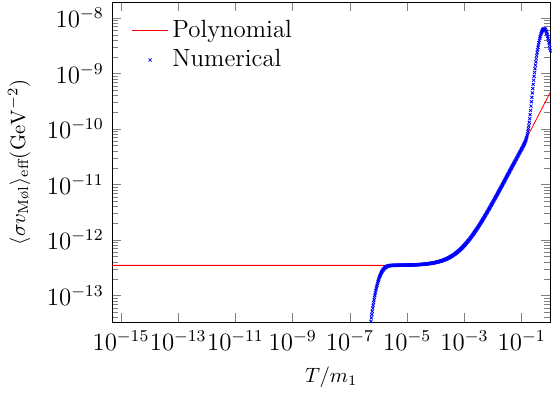}
\caption{Comparison between the results obtained for $\sveff$ by using the full expression \eqref{eqn:sigmavefffin2}
and the polynomial expansion at the first order by using \eqref{eqn:svexpansionlowT}. 
}
\label{fig:sveffnosplit}
\end{figure}

We can plug those expressions into $D$, $\mathcal{K}_n$ and $\sveff$, obtaining the result (to the 10th order):
\begin{align}
&\sveff = \frac{1}{4 m_1^2} \Bigg\{  W_0 + \frac{1}{x}\qty(-3 W_0 +\frac{3 W_1}{2}) + \frac{1}{x^2}\qty(6 W_0-3 W_1+\frac{15 W_2}{8})+\nonumber \\
&\, +\frac{1}{x^3}\qty(-\frac{75 W_0}{8}+\frac{75 W_1}{16}-\frac{15 W_2}{16}+\frac{35 W_3}{16})+\nonumber \\%
&\, +\frac{1}{x^4}\qty(\frac{23445 W_0}{2048}-\frac{1485 W_1}{256}-\frac{1575 W_2}{256}-\frac{525 W_3}{64}+\frac{315 W_4}{128})+\nonumber\\
&\, +\frac{1}{x^5}\qty(-\frac{17505 W_0}{2048}+\frac{19395 W_1}{4096}+\frac{11925 W_2}{512}+\frac{9975 W_3}{512}-\frac{4725 W_4}{512}+\frac{693 W_5}{256})+\nonumber\\
&\, +\frac{1}{x^6}\left( -\frac{222885 W_0}{32768}+\frac{13095 W_1}{8192}-\frac{878175 W_2}{16384}+\right.\nonumber \\
&\left.\qquad\quad -\frac{74025 W_3}{2048}+\frac{89775 W_4}{4096}-\frac{10395 W_5}{1024}+\frac{3003 W_6}{1024} \right)\nonumber \\
&\, +\frac{1}{x^7}\left( \frac{1661715 W_0}{32768}-\frac{1264815 W_1}{65536}+\frac{3173175 W_2}{32768}+\frac{1800225 W_3}{32768}\right.\nonumber\\
&\left.\qquad\quad -\frac{666225 W_4}{16384}+\frac{197505 W_5}{8192}+\frac{6435 W_7}{2048}-\frac{45045 W_6}{4096}\right) +\nonumber \\
&\, +\frac{1}{x^8}\left(-\frac{1379496825 W_0}{8388608}+\frac{32645025 W_1}{524288}-\frac{76137975 W_2}{524288}-\frac{8594775 W_3}{131072}+\right.\nonumber\\
&\left.\qquad\quad +\frac{16202025 W_4}{262144}-\frac{1465695 W_5}{32768}+\frac{855855 W_6}{32768}+\frac{109395 W_8}{32768}-\frac{96525 W_7}{8192}\right)+\nonumber\\
&\, +\frac{1}{x^9}\left(-\frac{13671950879025 W_0}{4294967296}-\frac{2855855475 W_1}{16777216}+\frac{186553125 W_2}{1048576}+\frac{45808875 W_3}{1048576}+\right.\nonumber \\
&\left.\qquad\quad -\frac{77352975 W_4}{1048576}+\frac{35644455 W_5}{524288}-\frac{6351345 W_6}{131072}+\right.\nonumber \\
&\left.\qquad\quad +\frac{1833975 W_7}{65536}-\frac{1640925 W_8}{131072}+\frac{230945 W_9}{65536}\right)+ \nonumber \\
 &\, +\frac{1}{x^{10}}\left(-\frac{38822473644075 W_0}{8589934592}-\frac{43047242435475 W_1}{8589934592}-\frac{10584016875 W_2}{67108864}+\right.\nonumber\\
&\left.\qquad\quad +\frac{275065875 W_3}{4194304}+ \frac{412279875 W_4}{8388608}-\frac{170176545 W_5}{2097152}+\frac{154459305 W_6}{2097152}\right.\nonumber\\
&\left.\qquad\quad -\frac{13610025 W_7}{262144}+\frac{31177575 W_8}{1048576}-\frac{3464175 W_9}{262144}+\frac{969969 W_{10}}{262144}\right)\Bigg\}\,,\label{eqn:svexpansionlowT}%
\end{align}
\begin{table}
 \centering\
 \[
 \begin{array}{ccc}
 \toprule
  & \text{Symbolic expression} & \text{Value} \\
  \midrule
  \beta_{0} & 1 & 1\\
  \beta_{1} & -b_1 & -\dfrac{15}{4}\\[2ex]
  \beta_{2} & b_1^2-b_2 & \dfrac{285}{32}\\[2ex]
  \beta_{3} & -b_1^3+2 b_2 b_1-b_3 & -\dfrac{2115}{128}\\[2ex]
  \beta_{4} & b_1^4-3 b_2 b_1^2+2 b_3 b_1+b_2^2-b_4 & \dfrac{51435}{2048}\\[2ex]
  \bottomrule
 \end{array}
 \]
 \caption{Symbolical expressions and values for the coefficients $\beta_r$.}
 \label{tab:betar}
\end{table}
which correctly reproduces the results in \cite{Srednicki:1988ce} and \cite{Cannoni:2013bza}.
From a numerical perspective, the error on $W_n$ for $n\geq 2$ will be large. Therefore, it is recommended to stop at the order 1 or 2. 
We show the results obtained for the pMSSM\footnote{The phenomenological Minimal Supersymmetric extension 
of the Standard Model.}
in \DarkPack in a scenario where the dark matter candidate has 
a mass $m_1 \approx \SI{200}{\giga\electronvolt}$ in \autoref{fig:sveffnosplit}.
We see that truncating at the first order can give a very satisfactory result, since the resulting curves for $\sveff(T)$ computed respectively with the numerical evaluation of Eq.~\eqref{eqn:sigmavefffin2} and of the asymptotic behaviour \eqref{eqn:svexpansionlowT} are compatible.
Moreover, from the figure, one can notice that the numerical implementation of the full formula for $\sigmav$ 
fails to deliver reliable results for $T \lesssim 10^{-6}m_1$.

\section{Case of particles with small mass splitting}\label{ssec:lowtbodysplit}
The result shown in \autoref{ssec:lowtbody} is correct up to a defined order, under the hypothesis that 
there are no new physics species with a mass close to the one of the dark matter candidate.\footnote{Except, of course, 
the candidate itself.}
In models with a $\mathbb{Z}_2$ symmetry, such a particle is the lightest of a set. Let us suppose that
there are $M\leq N$ particles nearly degenerate in mass. In such a case, we need to retain their 
contributions to the denominator in \eqref{eqn:sigmavefffin2}. Let us define:
\begin{align}
 x &= \frac{m_1}{T}, & x_i &= \frac{m_i}{T}, & c_i &= \frac{g_i}{g_1}\frac{m_i^2}{m_1^2}, & \Delta x_i &= x_i -x,
\end{align}
for $i = 1, \dots, M$. We can perform the same change of variable as in \autoref{ssec:lowtbody} which leads to:
\begin{align}
 \sveff &= \frac{x}{2m_1^2} \frac{1}{\qty[\sum_{i=1}^M c_i K_2\qty(\frac{m_i}{T})]^2}\int_1^{+\infty} \dd{y} \sqrt{y-1}\Weff(y)K_1\qty(2x\sqrt{y}).
\end{align}
After expanding $\Weff$ around $y=1$ as in the previous case we obtain:
\begin{align}
\sveff = 
\sum\limits_{n=0}^{\infty} \frac{1}{n!} W_n
\mathcal{H}_n (x).
\label{eqn:svexpansiondiff}
\end{align}
where we have defined:
\begin{align}
\mathcal{H}_n (x)&=\frac{x}{2 m_1^2} \frac{1}{\qty[\sum_{i=1}^M c_i K_2\qty(\frac{m_i}{T})]^2}I_n (x),
\end{align}
and where $I_n$ is the same as the one defined in \eqref{eqn:kandI}. 
Note that $\mathcal{H}_n$ and $\mathcal{K}_n$ differ only for the Bessel functions in the denominator. 
With some manipulations, we can treat $\mathcal{H}_n$ similarly as done with $\mathcal{K}_n$.
In fact, we already know how to write $I_n$ as Laurent series.

Let us define the quantity:
\begin{align}
 D(x) &= \qty[\sum_{i=1}^M c_i K_2\qty(x_i)]^2.
\end{align}
Our goal is to expand $D(x)$ at its first order in $\Delta x_i$ and at an arbitrary order in $x$.
Let us expand the square:
\begin{align}
 D(x) &= \qty[\sum_{i=1}^M c_i K_2\qty(x + \Delta x_i)]^2\nonumber \\
      &= 2\sum_{i=1}^M c_i K_2(x + \Delta x_i) \sum_{j=i}^M \qty(1 - \frac{\delta_{ij}}{2})c_j K_2(x + \Delta x_j).
\end{align}
At the first order in $\Delta x_i$ we have also:
\begin{align}\label{eqn:expansionk2dx}
 K_2\qty(x + \Delta x_i) = K_2(x) + K_2'(x)\Delta x_i,
\end{align}
where $K_2'(x)=\dv{K_2}{x} {(x)}$. Therefore:
\begin{align}
 D(x) &= 2\sum_{i=1}^M \sum_{j=i}^M c_i \qty(1 - \frac{\delta_{ij}}{2})c_j  \qty(K_2(x) + K_2'(x) \Delta x_i) \qty(K_2(x) + K_2'(x) \Delta x_j) \nonumber\\
      &=  2\sum_{i=1}^M \sum_{j=i}^M c_i \qty(1 - \frac{\delta_{ij}}{2})c_j  \qty(K_2^2(x) +K_2(x)K_2'(x)\qty(\Delta x_i +\Delta x_j)+o\qty(\Delta x^2)).
\end{align}
Separating the constant terms from the linear terms in $\Delta x_i$, and using the identity $2K_2(x) K_2'(x) = K_2^2(x)$ we obtain:
\begin{align}
 D(x) &= \qty[\sum_{i=1}^M \sum_{j=i}^M c_i \qty(2 - \delta_{ij})c_j ]K_2^2(x) + \qty[\sum_{i=1}^M \sum_{j=i}^M c_i \qty(1 - \frac{\delta_{ij}}{2})c_j \qty(\Delta x_i +\Delta x_j)] \qty(K_2^2)'(x),
\end{align}
where $\qty(K_2^2)'(x) = \dv{K_2^2}{x}{(x)}$. Let us define:
\begin{align}\label{eqn:rhoetatilde}
 \tilde \rho &= \qty[\sum_{i=1}^M \sum_{j=i}^M c_i \qty(2 - \delta_{ij})c_j ],  & \tilde \eta =  -\qty[\sum_{i=1}^M \sum_{j=i}^M c_i \qty(1 - \frac{\delta_{ij}}{2})c_j \qty(\Delta x_i +\Delta x_j)],
\end{align}
and use the expansions \eqref{eqn:asympK22n} and \eqref{eqn:k22prime}:
\begin{align}
 D(x) &= \frac{\pi}{2x} \ee^{-2x} \qty[ \tilde \rho (1 + B(x)) + \tilde \eta ( 2 + \tilde B(x))] \nonumber\\
      &=  \frac{\pi}{2x} \ee^{-2x} \qty[ \tilde \rho + 2\tilde \eta + \tilde \rho B(x) + \tilde \eta \tilde B(x)]\nonumber \\
      &= \frac{\pi}{2x} \ee^{-2x} \gamma \qty[1 + \rho B(x) + \eta \tilde B(x)],
\end{align}
with 
\begin{align}
 \gamma &=\tilde \rho + 2\tilde \eta, & \rho &= \frac{\tilde \rho}{\gamma}, & \eta &= \frac{\tilde \eta}{\gamma}.\label{eqn:rhoeta}
\end{align}
We can therefore write:
\begin{align}\label{eqn:1overDdef}
 \frac{1}{D(x)} = \frac{2x}{\pi \gamma} \ee^{2x}\frac{1}{1 + F(x)},
\end{align}
with:
\begin{align}\label{eqn:defF}
 F(x) =  \rho B(x) + \eta \tilde B(x).
\end{align}
The asymptotic behaviour at large $x$ for $F$ is proportional to $1/x$, since for both $B$ and $\tilde B$ it is proportional to $1/x$.
This means that we can treat \eqref{eqn:1overDdef} with the geometric expansion
\begin{align}
 \frac{1}{D(x)} = \frac{2x}{\pi \gamma} \ee^{2x}\sum_{r=0}^\infty (-1)^r F^r(x).
\end{align}
At this point we have found the same form as in the previous case, and we can treat it similarly.
Moreover, we have chosen to use the definition \eqref{eqn:defF} for $F$, because it has the advantage of 
being straightforward to reduce to the order 0 in $\Delta x_i$, which is the case if more particles 
have exactly the same mass. In fact, $\tilde B$ is from the first order of the expansion of 
$K_2^2(x+\Delta x_i)$ and also for $\Delta x_i =0$ we have $\eta =0$.

Thus, apart from a factor $1/\gamma$ and the replacement $B\to F$, we have found the same expression as 
in the previous case:
\begin{align}\label{eqn:sveffseriesdiff1}
 \sveff &= \frac{1}{2 \sqrt{\pi} m_1^2\gamma}
{x^2}\sum_{r=0}^{\infty} (-1)^r F^r(x) \sum\limits_{n=0}^{\infty} \qty[\frac{1}{n!} W_n\Gamma\qty(n+\frac{3}{2})\sum_{p = n+2}^\infty g_{n,p}x^{-p}].
\end{align}
This does not change the orders to which we need to truncate the series 
once we know that we want the result to a given $N\ped{max}$. By using the results in \eqref{eqn:ordersexpansion}:
\begin{align}\label{eqn:ordersexpansion1}
 0 \leq &r \leq N\ped{max}, & 0 \leq &n \leq N\ped{max}, & n+2 \leq &p \leq N\ped{max} +2.
\end{align}
The difference is that here the coefficients of $F$ up to a given order depend on model-dependent 
quantities, i.e.~$\rho$ and $\eta$.
We can use the same procedure as before, since we know that the geometric sum in $F$ has to be truncated 
at the order $N\ped{max}$. 
Hence, it is enough to expand each power $F^r$ separately as a function of $B$ and 
$\tilde B$, for which we know the coefficients, at the first order in $\eta$.\footnote{Since we 
are treating the first order in $\Delta x_i$.}
We automated this calculation in the \texttt{Wolfram Mathematica} notebook in the ancillary files.
Parametrising the geometric expansion as:
\begin{align}
 \frac{1}{1+F(x)} =& \sum_{r = 0}^\infty \frac{\phi_r}{x^r},
 \end{align}
we have, for $N\ped{max} = 4$, the following expressions:
\begin{subequations}
\begin{align}
\phi_0 =& 1,\\
\phi_1 =& -\eta  \tilde{b}_1-\rho  b_1,\\
\phi_2 =& \eta  \left(2 \rho  b_1 \tilde{b}_1-\tilde{b}_2\right)+\rho ^2 b_1^2-\rho  b_2,\\
\phi_3 =& \eta  \left(-3 \rho ^2 b_1^2 \tilde{b}_1+2 \rho  b_1 \tilde{b}_2+2 \rho  b_2 \tilde{b}_1-\tilde{b}_3\right)-\rho ^3 b_1^3+2 \rho ^2 b_2 b_1-\rho  b_3,\\
\phi_4 =& \eta  \left(4 \rho ^3 b_1^3 \tilde{b}_1-3 \rho ^2 b_1^2 \tilde{b}_2-6 \rho ^2 b_2 b_1 \tilde{b}_1+2 \rho  b_1 \tilde{b}_3+2 \rho  b_3 \tilde{b}_1+2 \rho  b_2 \tilde{b}_2-\tilde{b}_4\right)\nonumber\\
&+\rho ^4 b_1^4-3 \rho ^3 b_2 b_1^2+2 \rho ^2 b_3 b_1+\rho ^2 b_2^2-\rho  b_4.
\end{align}
\end{subequations}
The values of the $b_i$'s and $\tilde b_i$'s can be found in \autoref{tab:abbtilde}.

Plugging into the expression of $\sveff$, and replacing the $g_{n,p}$ with their values,\footnote{This simplifies dramatically the expressions, since many of them vanish.} we obtain:
\begin{align}
 \sveff = \frac{1}{{4 \gamma m_1^2}}\Bigg\{&W_0\phi_0 +\dfrac{1}{x} \Bigg[
W_0 \left(\frac{3 \phi _0}{4}+\phi _1\right)
+ \frac{3 W_1 \phi _0}{2}
\Bigg]\nonumber \\
+\dfrac{1}{x^{2}} \Bigg[&
W_0 \left(-\frac{3 \phi _0}{32}+\frac{3 \phi _1}{4}+\phi _2\right)
+ W_1 \left(\frac{21 \phi _0}{8}+\frac{3 \phi _1}{2}\right)
+ \frac{15 W_2 \phi _0}{8}
\Bigg]\nonumber \\
+\dfrac{1}{x^{3}} \Bigg[&
W_0\phi_0 \left(\frac{15 \phi _0}{128}-\frac{3 \phi _1}{32}+\frac{3 \phi _2}{4}+\phi _3\right)
+ W_1 \left(\frac{75 \phi _0}{64}+\frac{21 \phi _1}{8}+\frac{3 \phi _2}{2}\right)\nonumber \\
&+ W_2 \left(\frac{195 \phi _0}{32}+\frac{15 \phi _1}{8}\right)
+ \frac{35 W_3 \phi _0}{16}
\Bigg]\nonumber \\
+\dfrac{1}{x^{4}} \Bigg[&
W_0 \left(\frac{15 \phi _1}{128}-\frac{3 \phi _2}{32}+\frac{3 \phi _3}{4}+\phi _4\right)
+ W_1 \left(\frac{75 \phi _1}{64}+\frac{21 \phi _2}{8}+\frac{3 \phi _3}{2}\right)\nonumber \\
&+ W_2 \left(\frac{195 \phi _1}{32}+\frac{15 \phi _2}{8}\right)
+ \frac{35 W_3 \phi _1}{16}
+ \frac{315 W_4 \phi _0}{128}
\Bigg]\Bigg\}.\label{eqn:svspurious}
\end{align}
From this expression one can check that in the previous hypothesis (i.e.~$\gamma = \rho=1$ and $\eta=0$) 
we recover the same coefficients as in \eqref{eqn:svexpansionlowT}.
This expression, however, is correctly expanded until the order 4 in $1/x$, but it contains some spurious terms 
of higher orders in $\Delta x_i$. In order to eliminate them and consistently truncate at the first order, we have to 
do some more manipulations. Recalling the definitions \eqref{eqn:rhoeta} and \eqref{eqn:rhoetatilde}:
\begin{align}
 \rho &= \frac{1}{1 + 2\frac{\tilde \eta}{\tilde \rho}}, & \eta &= \frac{\tilde \eta}{\tilde \rho + 2\tilde \eta},
\end{align}
and now we can identify $\tilde \eta$ as the expansion parameter. At the first order we have:
\begin{align}
 \gamma^{-1} &= \frac{1}{\tilde\rho} \qty( 1 - 2\frac{\tilde\eta}{\tilde \rho}), &
 \rho^n &= 1 - 2n\frac{\tilde\eta}{\tilde \rho}, &
 \eta &= \frac{\tilde\eta}{\tilde \rho}. \label{eqn:1stordereta}
\end{align}
As we knew already, $\eta \sim \Delta x_i$, justifying the truncation of higher powers of $\eta$.
Since $\tilde\eta$ is always divided by $\tilde\rho$, we can choose $\eta$ as the expansion 
parameter.

\begin{figure}\centering
\includegraphics{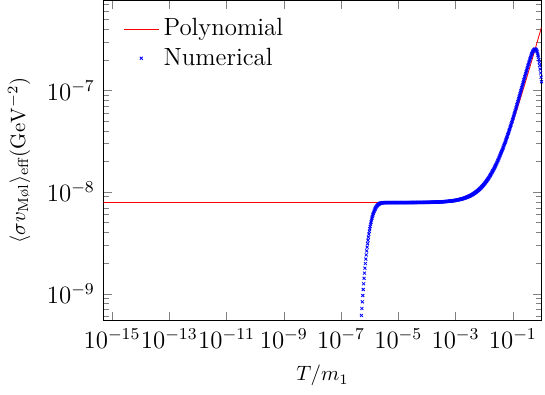}
\caption{Comparison between the results obtained for $\sveff$ using the full expression \eqref{eqn:sigmavefffin2}
and the polynomial expansion at the first order by using \eqref{eqn:svdiffcorrect}. 
Plot obtained by using \DarkPack \cite{Palmiotto:2022rvw}.
}
\label{fig:sveffdegenmass}
\end{figure}

In the formula \eqref{eqn:svspurious}, we can replace $\gamma$ with $\tilde\rho$ and $\phi_n$ with:
\begin{align}
 \tilde\phi_n = \qty(1 - 2\eta) \phi_n.
\end{align}
We are left with expressing $ \tilde\phi_n$ at the first order in $\eta$:
\begin{subequations}\label{eqn:phitildes}
\begin{align}
\tilde \phi_0 =& 1-2 \eta,\\
\tilde \phi_1 =& \eta  \left(3 b_1-\tilde{b}_1\right)-b_1,\\
\tilde \phi_2 =& \eta  \left(2 b_1 \tilde{b}_1-\tilde{b}_2-4 b_1^2+3 b_2\right)+b_1^2-b_2,\\
\tilde \phi_3 =& \eta  \left(-3 b_1^2 \tilde{b}_1+2 b_1 \tilde{b}_2+2 b_2 \tilde{b}_1-\tilde{b}_3+5 b_1^3-8 b_2 b_1+3 b_3\right)-b_1^3+2 b_2 b_1-b_3,\\
\tilde \phi_4 =& \eta  \left(4 b_1^3 \tilde{b}_1-3 b_1^2 \tilde{b}_2-6 b_2 b_1 \tilde{b}_1+2 b_1 \tilde{b}_3+2 b_3 \tilde{b}_1+2 b_2 \tilde{b}_2-\tilde{b}_4-6 b_1^4\right.\nonumber\\
          &\left.+15 b_2 b_1^2-8 b_3 b_1-4 b_2^2+3 b_4\right)+b_1^4-3 b_2 b_1^2+2 b_3 b_1+b_2^2-b_4.
\end{align}
\end{subequations}
The expression for $\sveff$ at the first order in $\Delta x_i$ and at the 4th order in $x^{-1}$ reads then:
\begin{align}
 \sveff = \frac{1}{{4 \tilde \rho m_1^2}}\Bigg\{&W_0\tilde \phi_0 +\dfrac{1}{x} \Bigg[
W_0 \left(\frac{3 \tilde \phi _0}{4}+\tilde \phi _1\right)
+ \frac{3 W_1 \tilde \phi _0}{2}
\Bigg] \nonumber \\
+\dfrac{1}{x^{2}} \Bigg[&
W_0 \left(-\frac{3 \tilde \phi _0}{32}+\frac{3 \tilde \phi _1}{4}+\tilde \phi _2\right)
+ W_1 \left(\frac{21 \tilde \phi _0}{8}+\frac{3 \tilde \phi _1}{2}\right)
+ \frac{15 W_2 \tilde \phi _0}{8}
\Bigg] \nonumber \\
+\dfrac{1}{x^{3}} \Bigg[&
W_0 \left(\frac{15 \tilde \phi _0}{128}-\frac{3 \tilde \phi _1}{32}+\frac{3 \tilde \phi _2}{4}+\tilde \phi _3\right)
+ W_1 \left(\frac{75 \tilde \phi _0}{64}+\frac{21 \tilde \phi _1}{8}+\frac{3 \tilde \phi _2}{2}\right)\nonumber \\
&+ W_2 \left(\frac{195 \tilde \phi _0}{32}+\frac{15 \tilde \phi _1}{8}\right)
+ \frac{35 W_3 \tilde \phi _0}{16}
\Bigg] \nonumber \\
+\dfrac{1}{x^{4}} \Bigg[&
W_0 \left(\frac{15 \tilde \phi _1}{128}-\frac{3 \tilde \phi _2}{32}+\frac{3 \tilde \phi _3}{4}+\tilde \phi _4\right)
+ W_1 \left(\frac{75 \tilde \phi _1}{64}+\frac{21 \tilde \phi _2}{8}+\frac{3 \tilde \phi _3}{2}\right)\nonumber \\
&+ W_2 \left(\frac{195 \tilde \phi _1}{32}+\frac{15 \tilde \phi _2}{8}\right)
+ \frac{35 W_3 \tilde \phi _1}{16}
+ \frac{315 W_4 \tilde \phi _0}{128}
\Bigg]\Bigg\},\label{eqn:svdiffcorrect}
\end{align}
with $\tilde \rho$ defined in \eqref{eqn:rhoetatilde}, $\eta$ defined in \eqref{eqn:1stordereta} and $\tilde \phi_n$s 
defined in \eqref{eqn:phitildes}. 
Note that we showed the results at the order 4 since the expression is already very complicated, but the 
\texttt{Wolfram Mathematica} notebook in the ancillary files  allows us to obtain the correct result at any given order.
We would like to comment the result, by noticing that, in this form, the coefficients of $x^{-n}$ depend on $T$, but 
they are linear in $1/T$, therefore, there is a term proportional to $\eta \sim 1/T$ in $\tilde \phi_0$. 
However, we recall that the requirement for the mass degeneracy to contribute is to have 
$\Delta x_i$ small enough to allow the expansion \eqref{eqn:expansionk2dx}.
At very low temperatures, only the truly degenerate species contribute, and for $\Delta x_i=0$ we have $\eta =0$,
so the 0th order contribution does not have a divergence, and the only deviation from the previous formula 
\eqref{eqn:svexpansionlowT} is contained in the $\tilde \rho$ in the prefactor. 
In \autoref{fig:sveffdegenmass} we show the results for the MSSM, in which we considered the three lightest neutralinos 
to have the same mass of $\SI{200}{\giga\electronvolt}$. 

\begin{table}
 \centering
 \begin{tabular}{*{6}{>{$}c<{$}}}
 \toprule 
 n & 0 & 1 & 2 & 3 & 4 \\
 \midrule
 \beta_n & 1	&  -\dfrac{15}{4}	&\dfrac{285}{32} & -\dfrac{2115}{128} & \dfrac{51435}{2048}\\[2ex]
 \midrule
 \lambda_n &   -2   & -\dfrac{11}{4} & \dfrac{165}{32} & -\dfrac{5295}{128} & \dfrac{297135}{2048}\\[2ex]
 \bottomrule
 \end{tabular}
\caption{Parameters of $\tilde\phi_n$ in the form $\tilde\phi_n = \beta_n + \eta \lambda_n$.}
\label{tab:betaphitilde}
\end{table}

\section{Conclusion}\label{ssec:conclusion}

We showed in detail how to derive the formula for the expansion of $\sveff$ at low temperatures
at an arbitrary order in $x=m_1/T$ and at the first order in the mass splittings $x_i -x$, 
providing also two \texttt{Wolfram Mathematica} notebooks that allow us to perform both the expansions
at an arbitrary order in $x$.

The implementation of formula \eqref{eqn:svexpansionlowT} in the software \DarkPack has shown
the necessity of the usage of such an expansion at low temperatures, as the implementation 
of the full formula \eqref{eqn:sigmavefffin2} fails to provide a numerically stable result.
The result is publicly available in \DarkPack \texttt{1.2}.

As a continuation of this work, we will implement also the formula with mass degeneracy in \DarkPack, 
allowing to have a more reliable tool for the computation of dark matter densities, especially 
in models with low freeze-out temperatures. 

The overall improved stability of the algorithm in \DarkPack will also be used to solve a system 
of Boltzmann equations - one for each species at its own temperature $T_i$ - to be able to study more general scenarios, in which all the particles of the same species are in thermal 
equilibrium between them, but not necessarily with the particles of other species.
In particular, such hypotheses will allow us the study of freeze-in scenarios, or of models 
where there is more than one particle as dark matter candidate.

\cleardoublepage
\phantomsection
\appendix

\begin{comment}
\section{Bessel functions}
\subsection{Integral representation for positive variables}
Let us start by the Bessel functions $K_n$ with real and positive argument. According to \cite{abramowitz1968handbook},
for $z>0$ and $n$ natural, $K_n(z)$ is 
\begin{align}\label{eqn:intergalKn}
 K_n(z) = \frac{\sqrt{\pi}z^n}{2^n \Gamma\qty(n + \frac{1}{2})} \int_1^{+\infty} \dd t (t^2 -1)^{n-\frac{1}{2}} \ee^{-zt}.
\end{align}
Using the property $\Gamma\qty(n + \frac{1}{2}) = \frac{(2n)!}{4^n n!}\sqrt{\pi}$, it yields
\begin{align}
 K_n(z) = \frac{2^n n!z^n}{(2n)!} \int_1^{+\infty} \dd t (t^2 -1)^{n-\frac{1}{2}} \ee^{-zt}.
\end{align}
Making the change of variable $t\to \tau /z$, the previous result becomes
\begin{align}
 K_n(z) = \frac{2^n n!}{2n(2n-2)!z^n} \int_z^{+\infty} \dd \tau (\tau^2 - z^2)^{n-\frac{3}{2}} \tau \ee^{-\tau}.
\end{align}
Integrating by parts:
\begin{align}
 K_n(z) = \frac{2^{n-1} (n-1)!}{(2n-2)!z^n} \int_z^{+\infty} \dd \tau (\tau^2 - z^2)^{n-\frac{3}{2}} \tau \ee^{-\tau}.
\end{align}
For $n=2$:
\begin{align}\label{eqn:K2forn}
 K_2(z) = \frac{1}{z^2} \int_z^{+\infty} \dd \tau \sqrt{\tau^2 - z^2} \tau \ee^{\tau}.
\end{align}
\end{comment}
\section{Asymptotic expansions of the Bessel functions}
\label{ssec:Bessel}
\noindent The asymptotic form of $K_n(z)$ for $\abs{z} \to + \infty$ and $\abs{\arg{z}} < 3\pi/2$ can be written as~\cite{abramowitz1968handbook}:
\begin{align}\label{eqn:asympKn}
 K_n(z) \sim \sqrt{\frac{\pi}{2z}} \ee^{-z} \qty[ 1 + \frac{\mu -1}{8z} + \frac{(\mu -1)(\mu -9)}{2!(8z)^2} +
        \frac{(\mu -1)(\mu -9)(\mu -25)}{3!(8z)^3} + o\qty(z^{-4})]
\end{align}
where $\mu = 4n^2$.

For $n=2$ we define the quantity $A(x)$ such that:
\begin{align}\label{eqn:asympK2n}
 K_2(x) \sim \sqrt{\frac{\pi}{2x}} \ee^{-x} \qty[ 1 + A(x)]\,,
\end{align}
and we write it in the form: 
\begin{align}
 A(x) = \sum_{m=1}^\infty a_m x^{-m}.
\end{align}
Analogously, we can define the quantity $B(x)$ such that:
\begin{align}\label{eqn:asympK22n}
 K_2^2(x) \sim \frac{\pi}{2x} \ee^{-2x} \qty[ 1 + B(x)]\,,
\end{align}
implying the relation $B = 2A + A^2$. 

Finally, it is helpful to write the asymptotic form of the first derivative of $K_2^2$ as:
\begin{align}\label{eqn:k22prime}
 \qty(K_2^2)'(x) &\sim - \frac{\pi}{2x} \ee^{-2x}\qty[2 + \tilde B(x)]
\end{align}
with: 
\begin{align}
 \tilde B(x) &= 2B(x) + \frac{1}{x} + \frac{B(x)}{x} - B'(x)\,.
\end{align}
where the $'$ denotes the derivative with respect to $x$.

\begin{table}
\[
 \begin{array}{|c||c|c|c|}
\hline
  m & a_m & b_m & \tilde b_m \\
\hline\hline

1 & \dfrac{15}{8} & \dfrac{15}{4} & \dfrac{17}{2}\\[2ex]

2 & \dfrac{105}{128} & \dfrac{165}{32} & \dfrac{285}{16}\\[2ex]

3 & -\dfrac{315}{1024} & \dfrac{315}{128} & \dfrac{1305}{64}\\[2ex]

4 & \dfrac{10395}{32768} & \dfrac{315}{2048} & \dfrac{10395}{1024}\\[2ex]

5 & -\dfrac{135135}{262144} & -\dfrac{2835}{8192} & \dfrac{315}{4096}\\[2ex]

6 & \dfrac{4729725}{4194304} & \dfrac{61425}{65536} & -\dfrac{6615}{32768}\\[2ex]

7 & -\dfrac{103378275}{33554432} & -\dfrac{779625}{262144} & \dfrac{80325}{131072}\\[2ex]

8 & \dfrac{21606059475}{2147483648} & \dfrac{90904275}{8388608} & -\dfrac{8887725}{4194304}\\[2ex]

9 & -\dfrac{655383804075}{17179869184} & -\dfrac{1497971475}{33554432} & \dfrac{138305475}{16777216}\\[2ex]

10 & \dfrac{45221482481175}{274877906944} & \dfrac{55124944875}{268435456} & -\dfrac{4793914125}{134217728}\\[2ex]
\hline
 \end{array}
\]
\caption{Coefficients of the Laurent series in the asymptotic expansion of $K_2$, $K_2^2$ and $\qty(K_2^2)'$.}
\label{tab:abbtilde}
\end{table}

Note that since $A \sim 1/x$,\footnote{by neglecting a non-null real factor.} we have $B\sim 1/x$ and $\tilde B \sim 1/x$:
\begin{align}\label{eqn:BBtildeseries}
 B(x) &= \sum_{m=1}^\infty b_m x^{-m}, & \tilde B(x) &= \sum_{m=1}^\infty \tilde b_m x^{-m}.
\end{align}
Therefore, the coefficients can be determined using the relations 
\begin{align}
 b_m &= 2a_m + \sum_{k=1}^{m} a_k a_{m-k}, & & & \forall m \geq 1\\
 \tilde b_1 &= 2b_1 +1, & \tilde b_m &= 2b_m +m b_{m-1}. & \forall m \geq 2
\end{align}
The coefficients $a_m$ are known, and their values are given in Table~\ref{tab:abbtilde}, thus we can compute all the other parameters.

\section{Meijer functions}
\label{ssec:Meijer}

The Meijer functions are defined as:\footnote{See {\it e.g.}~definition (9.301) in Ref.~\cite{GradshteynRyzhik2007}.}
\begin{equation}
G^{m,n}_{p,q}\left(z\Bigg\rvert
\begin{array}{c}
 a_1,...,a_n,a_{n+1},...,a_p\\
b_1,...,b_m,b_{m+1}...,b_q
\end{array}
\right)
=
\frac{1}{2\pi\ii}\int \dd{s} \dfrac{\prod_{j=1}^m \Gamma(b_j-s) \prod_{j=1}^n \Gamma(1-a_j+s)}{\prod_{j=m+1}^q\Gamma(1-b_j+s) \prod_{j=n+1}^p \Gamma(a_j-s)}x^s
\end{equation}
where $0\leq m \leq q$, $0\leq n\leq p$, and the poles of $\Gamma(b_j-s)$ must not coincide with the poles of $\Gamma(1-a_j+s)$ for any pair $(j,k)$
with $1\leq j \leq n$, $1 \leq k \leq m$.

The following property holds:\footnote{See {\it e.g.}~equation (6.592.4) in Ref.~\cite{GradshteynRyzhik2007}.}
\begin{flalign}
\int_{1}^{\infty} \dd x x^{\lambda} (x-1)^{\mu-1} K_\nu (a\sqrt{x})=
\Gamma(\mu) 2^{2\lambda-1} a^{-2\lambda} G_{1,3}^{3,0}
\begin{pmatrix}
\frac{a^2}{4} \Biggr\rvert
\begin{array}{c}
 0 \\
-\mu, \frac{\nu}{2}+\lambda,-\frac{\nu}{2}+\lambda
\end{array}
\end{pmatrix}.
\label{I_GR}
\end{flalign}
For $\lambda=0$, $\mu=3/2+n$ and $\nu=1$, the asymptotic form of the $G$-function at the right hand side is: 
\begin{align}\label{eqn:Gnexpansion}
 G_{1,3}^{3,0}
\begin{pmatrix}
x^2 \Biggr\rvert
\begin{array}{c}
 0 \\
 -\frac{1}{2},\frac{1}{2},-n-\frac{3}{2}
\end{array}
\end{pmatrix} 
=\sqrt{\pi}\ee^{-2x} G_n(x),
\end{align}
where $G_n(x)$ is a generalized series:
\begin{align}
 G_n(x) = \sum_{p = n+2}^\infty g_{n,p}x^{-p}.
\end{align}
The results of the expansion of $G_n$ for $0 \leq n \leq 10$ to the 12th order are:
\begin{subequations}\label{eqn:gnpvalues}
\begin{align}
 G_0 &= \frac{1}{x^2}+\frac{3}{4 x^3}-\frac{3}{32 x^4}+\frac{15}{128 x^5}, \\
 G_1 &= \frac{25}{32 x^5}+\frac{7}{4 x^4}+\frac{1}{x^3}, \\
 G_2 &= \frac{13}{4 x^5}+\frac{1}{x^4}, \\
 G_\nu &= \frac{1}{x^{\nu+2}}, &\forall \, 3 \leq \nu \leq 10.
\end{align}
\end{subequations}

\bibliography{database}

\end{document}